# Electric field modulation of the non-linear areal magnetic anisotropy energy

Yong-Chang Lau[1,2], Peng Sheng[1], Seiji Mitani[1], Daichi Chiba[3] and Masamitsu Hayashi[1,2]

[1]*National Institute for Materials Science, Tsukuba 305-0047, Japan*

[2]*Department of Physics, The University of Tokyo, Bunkyo, Tokyo 113-0033, Japan*

[3]*Department of Applied Physics, The University of Tokyo, Bunkyo, Tokyo 113-0033, Japan*

We study the ferromagnetic layer thickness dependence of the voltage-controlled magnetic anisotropy (VCMA) in gated CoFeB/MgO heterostructures with heavy metal underlayers. When the effective CoFeB thickness is below ~1 nm, the VCMA efficiency of Ta/CoFeB/MgO heterostructures considerably decreases with decreasing CoFeB thickness. We find that a high order phenomenological term used to describe the thickness dependence of the areal magnetic anisotropy energy can also account for the change in the areal VCMA efficiency. In this structure, the higher order term competes against the common interfacial VCMA, thereby reducing the efficiency at lower CoFeB thickness. The areal VCMA efficiency does not saturate even when the effective CoFeB thickness exceeds ~1 nm. We consider the higher order term is related to the strain that develops at the CoFeB/MgO interface: as the average strain of the CoFeB layer changes with its thickness, the electronic structure of the CoFeB/MgO interface varies leading to changes in areal magnetic anisotropy energy and VCMA efficiency.

Magnetization manipulation by means of electric field[1-5] is an attractive alternative to current-controlled schemes owing to its potential for realizing magnetic tunnel junctions (MTJ) with orders of magnitude reduction in energy consumption. However, due to the screening by the free electrons,[6] electric field effects are very often inefficient in metallic systems. Nowadays, many modern MTJ stacks incorporate Ta/ultrathin CoFe(B)/MgO based heterostructures, which exhibit a unique combination of high tunneling magnetoresistance for data readout and strong perpendicular magnetic anisotropy (PMA) for high-density non-volatile data retention as well as low spin-transfer torque switching current density.[7,8] Interestingly, in view of the interfacial origin of the PMA due to the hybridization of the oxygen *p*-orbitals and the iron *d*-orbitals at the ferromagnetic metal (FM)/oxide interface,[9,10] appreciable voltage-controlled magnetic anisotropy (VCMA)[11,12] and voltage-driven/assisted switching[13-16] have also been demonstrated in these structures.

The PMA modulation in such structures has been mainly attributed to the electric-field-induced charge accumulation/depletion at the FM/oxide interface, which subsequently influences the *p-d* orbitals hybridization and hence the PMA.[17] Meanwhile, other possible contributions including voltage-induced oxygen migration,[18] strain effect[19,20] and piezoelectricity[21] have been proposed. The areal efficiency of the VCMA is commonly evaluated by $\xi = \Delta K_{\text{eff}} \cdot t_{\text{eff}}/E$ where $\Delta K_{\text{eff}}$ is the change of the effective magnetic anisotropy energy (MAE) density due to an applied electric field $E$ and $t_{\text{eff}}$ is the effective FM layer thickness excluding the magnetic dead layer. Previous reports have shown that the value and even the sign of $\xi$ seem to be sensitive to the choice of the materials (underlayer, ferromagnet and capping oxide)[22-26] as well as the deposition conditions.[27] Typically, $\xi$ of polycrystalline

Ta/CoFeB/MgO interface is of the order of a few nano-erg/(V.cm) (a few tens of fJ/(V.m)).[12,14,15,28] Here we compare the CoFeB thickness ($t_{eff}$) dependence of VCMA in two perpendicularly-magnetized heterostructures: Ta/CoFeB/MgO and W/CoFeB/MgO. We find that $\xi$ decreases rather monotonically with decreasing $t_{eff}$. Our result challenges the archetypal assumption that VCMA is a purely interfacial effect, which would imply a thickness-independent $\xi$.

Wedged CoFeB-based heterostructures with different heavy metal (HM) underlayers, Sub./1 Ta/$t$ Co$_{20}$Fe$_{60}$B$_{20}$/2 MgO/1 Ta and Sub./3 W/$t$ Co$_{20}$Fe$_{60}$B$_{20}$/2 MgO/1 Ta (nominal thicknesses in nanometer), are grown on 10 x 10 mm$^2$ thermally oxidized Si substrates by magnetron sputtering at ambient temperature. In addition, uniform films with fixed CoFeB thicknesses are grown under similar conditions for vibrating sample magnetometry (VSM). Prior to the patterning process, samples are annealed in vacuum at 300$^o$C for one hour without the application of an external field. Hall bars having a channel width of ~35 μm and a distance of ~45 μm between probes measuring longitudinal resistance $R_{xx}$ are fabricated using standard optical lithography and Ar ion milling. 50 nm-thick HfO$_2$ are grown by atomic layer deposition on patterned structures as the main gate oxide followed by dry etching. Finally, 5 Ta/100 Au bilayer is deposited for top contact.

Figure **1**(a) shows a cross-sectional schematic view of the device structure and the coordinate system used in this study. A gate voltage $V_g$ up to ± 14 V is applied between the top metal gate and the ground of the Hall bar, which generates an electric field along $z$, across the insulating 50 HfO$_2$/~1 TaO$_x$/2 MgO trilayer. By convention, a positive (negative) $V_g$ will induce

electron accumulation (depletion) at the CoFeB/MgO interface and we estimate the corresponding electric field at that interface due to $V_g$ = 14 V is around 0.26 V/nm. At various $V_g$, the Hall resistance ($R_{xy}$) of the devices is monitored while sweeping the external field. Measurements are performed in two configurations with the CoFeB magnetization pointing along +z (+$M_z$) and –z (-$M_z$). In order to minimize the current-induced spin-orbit effective fields[29,30], the source current is fixed at 0.1 mA throughout the study, which corresponds to a maximum current density of 3 x 10$^5$ A/cm$^2$.

We first focus on the magnetic properties of the annealed unpatterned films. The effective MAE density, $K_{eff}$ of a magnetic thin film is commonly expressed as the sum of a bulk, volumic contribution $K_B$, a demagnetizing term $-2\pi M_s^2$ and an interfacial term $K_i/t_{eff}$:[31-33]

$$K_{eff} = K_B - 2\pi M_s^2 + \frac{K_i}{t_{eff}} \quad (1)$$

$t_{eff} = t - t_{DL}$ is the effective FM layer thickness, obtained by subtracting the magnetic dead layer $t_{DL}$ from the nominal FM thickness, $t$. Figure **1**(b) shows the magnetic moment per unit area ($M/A$) as a function of the CoFeB nominal thickness $t$, for stacks with Ta and W underlayers. For both sample sets, a slope change near a common thickness of ~2 nm is observed, which is consistent with previous report.[32] The average saturation magnetization of CoFeB, $M_s$ as well as $t_{DL}$ are extracted from the slope and the x-intercept of the best linear fit to the data with thickness below 2 nm. We obtain $M_s$ = 1450 ± 100 emu/cm$^3$ and $t_{DL}$ = 0.39 ± 0.10 nm for the W underlayer films and $M_s$ = 1340 ± 60 emu/cm$^3$ and $t_{DL}$ = 0.40 ± 0.10 nm for the Ta underlayer films.

$K_{eff}$ can be evaluated from the area enclosed by the out-of-plane and the in-plane M-H loops at the first quadrant, with $K_{eff} > 0$ corresponds to PMA. It is convenient to plot the areal MAE density $K_{eff}t_{eff}$ as a function of $t_{eff}$, where the slope and the y-intercept at $t_{eff} = 0$ of the best linear fit represents respectively $K_B - 2\pi M_s^2$ and $K_i$, as shown in Figure **1**(c). Although the two data sets exhibit similar slope, their bulk anisotropy $K_B$ appears to be different, due to the higher $M_s$ of the samples with W underlayer. Substituting $M_s$ obtained from Figure **1**(b) yields $K_B$ = 2.9 ± 1.8 Merg/cm³ for CoFeB on W and $K_B$ = 0.6 ± 1.0 Merg/cm³ for CoFeB on Ta. We shall note that the uncertainty of $K_B$ can be large as it scales with the uncertainty of $M_s^2$. In terms of the interfacial anisotropy, Ta/CoFeB/MgO heterostructure exhibits a higher anisotropy ($K_i$ = 1.6 ± 0.2 erg/cm²) than that of W/CoFeB/MgO ($K_i$ = 1.3 ± 0.2 erg/cm²).[32,34] We highlight that for $t_{eff}$ < 1 nm, the experimental data deviate significantly from the ideal linear behavior.

Next, we study VCMA via the Hall resistance of patterned Hall bar devices exhibiting sufficient PMA (typically with an anisotropy field $H_k$ > 3 kOe). Assuming that the CoFeB layer acts as a macrospin, and in the limit where the polar tilting angle of the magnetization due to external field is small, the Hall resistance measured against an in-plane magnetic field, $R_{xy}(H)$, can be fitted by parabolic functions: $aH^2 + R_{xy}^0$. $R_{xy}^0$ denotes the Hall resistance in the absence of external magnetic field. $a$ can be expressed in terms of the amplitude of the anomalous Hall resistance ($\Delta R_{xy}^{AHE}$) as well as the in-plane anisotropy field $H_k$ of the CoFeB layer:[35,36]

$$a = \mp \frac{1}{4} \frac{\Delta R_{xy}^{AHE}}{H_k^2} \quad (2)$$

The ± sign corresponds to ±$M_z$. For a given $V_g$, $\Delta R_{xy}^{AHE} \equiv R_{xy}^0(+M_z) - R_{xy}^0(-M_z)$ is obtained by combining the data of ±$M_z$, as shown in the inset of Figure **2**(a). $H_k$ extracted from the Hall resistance of devices measured at $V_g$ = 0 V and those extracted from the VSM of unpatterned films are in reasonable agreement, as shown in the main panel of Figure **2**(a). For CoFeB grown on W underlayer, $H_k$ increases rather linearly with decreasing $t_\text{eff}$ and reaches ~ 12 kOe for $t_\text{eff}$ = 0.43 nm. In contrast, the $t_\text{eff}$ dependence of $H_k$ for heterostructures grown on Ta seed layer is clearly a concave function, with a maximum $H_k$ of ~ 6 kOe at $t_\text{eff}$ ~ 0.6 nm.

Typical change of $H_k$ as a function of the applied gate voltage $V_g$ is plotted in Figure **2**(b). Here, two devices with comparable CoFeB thickness ($t_\text{eff}$ ~ 0.8 nm) and anisotropy ($H_k$ ~ 6 kOe) but grown on different underlayers (Ta and W) are chosen for comparison. For both devices, $H_k$ is found to vary rather linearly with respect to $V_g$. Although we do find that the slope of $H_k$ vs. $V_g$ slightly changes at $V_g$ = 0, such change is much smaller than what has been reported in other systems.[23,37,38] Positive (negative) $V_g$, which results in charge accumulation (depletion) at the CoFeB/MgO interface, *decreases* (increases) PMA. The efficiency of the voltage-controlled $H_k$ is extracted from the slope of the best linear fit over the full $V_g$ range. Figure **2**(c) summarizes the underlayer material and the CoFeB effective thickness ($t_\text{eff}$) dependence of the $H_k$ slope change. CoFeB films grown on Ta show almost twice the sensitivity for a given $V_g$ modulation compared to that grown on W.

We estimate the areal efficiency of VCMA ($\xi$) assuming $K_i$ is the unique parameter that changes with $V_g$:

$$\xi = \frac{\Delta K_\text{eff} t_\text{eff}}{E} = \frac{\Delta K_\text{i}}{E} \approx \frac{1}{2}\frac{M_s t_\text{eff} \Delta H_k}{E} \quad (3)$$

$\xi$ as a function of $t_\text{eff}$ is plotted in Figure **3**. Surprisingly, $\xi$ exhibits strong dependence with varying $t_\text{eff}$. For Ta/CoFeB/MgO heterostructures, $\xi$ decreases in magnitude by nearly ~50% in the thickness range studied. We note that $\xi$ obtained from thicker CoFeB films approaches the literature value of polycrystalline Ta/CoFeB/MgO stacks with comparable CoFeB thickness.[12,14,15,28] However, such strong thickness dependence of $\xi$ has not been reported previously. A somewhat smaller reduction of $\xi$ (~21%) with decreasing $t_{eff}$ is also observed in W/CoFeB/MgO heterostructures.

We also note that for the thickness range over which the electric field effect was evaluated, the $K_{eff}t_{eff}$ versus $t_{eff}$ plot deviates significantly from the ideal linear behavior, in particular for the Ta underlayer films. Such non-linearity has been previously attributed to a thickness-dependent magneto-elastic anisotropy of general form $K_{elastic}(t_{eff}) = B_{eff}^{biaxial}(t_{eff})\epsilon(t_{eff})$,[31,39] where both the effective magneto-elastic coupling coefficient $B_{eff}^{biaxial}$ and the average in-plane biaxial strain $\epsilon$ depend on $t_{eff}$. $B_{eff}^{biaxial}$ of annealed Ta/CoFeB/MgO heterostructures was reported to be negative with its magnitude increasing with increasing $t_{eff}$. The $t_{eff}$ dependence of $\epsilon$ is understood by considering that a templated crystallization of CoFeB occurs at its interface with MgO. The degree of crystallization depends on the annealing temperature[40] and for the films studied here (annealed at ~300 °C) the CoFeB layer is predominantly amorphous but may contain small crystallites at the MgO interface.[41] The CoFeB can be strained wherein it is partially templated crystalized by the MgO layer while being essentially amorphous at the practically unstrained heavy metal/CoFeB interface. The asymmetry of the interfacial strain gives rise to a strain profile that may depend on $t_{eff}$.

Following Gowtham *et al.*,[39] $B_{eff}^{biaxial}$ and $\epsilon(t_{eff})$ can be modeled as

$$B_{eff}^{biaxial}(t_{eff}) \approx B_B + B_i/t_{eff} \quad (4)$$

$$\epsilon(t_\text{eff}) \approx \epsilon_0 + \gamma/t_\text{eff} \quad (5)$$

where $B_\text{B}$ and $B_\text{i}$ ($\epsilon_0$ and $\gamma$) are the bulk and interfacial contributions to the effective magneto-elastic coupling coefficient (strain). The resulting MAE density $K_\text{eff}$ including the $K_\text{elastic}$ contribution reads

$$K_\text{eff} = (K'_\text{B} - 2\pi M_\text{s}^2) + \frac{K'_\text{i}}{t_\text{eff}} + \frac{K_3}{t_\text{eff}^2} \quad (6)$$

with an effective bulk contribution $K'_\text{B} = K_\text{B} + B_\text{B}\epsilon_0$, an effective interface term $K'_\text{i} = K_\text{i} + B_\text{B}\gamma + B_\text{i}\epsilon_0$ and a new term $K_3 = B_\text{i}\gamma$ that scales with $1/t_\text{eff}^2$.

The phenomenological $K_3$ term is responsible for the reduction of $K_\text{eff}$ at lower thicknesses but has almost no effect when $K_\text{eff}t_\text{eff}$ versus $t_\text{eff}$ becomes linear ($t_\text{eff} \gtrsim 1.2$ nm). We fit the data shown in Figure **1**(c) using Eq. (6) with $(K'_\text{B} - 2\pi M_\text{s}^2)$, $K'_\text{i}$, and $K_3$ as fitting parameters. The results are summarized in Table 1 in comparison with the parameters obtained from the fitting without the $K_3$ term (i.e. fitting using Eq. (1)). We find that the interfacial contribution ($K'_\text{i}$) to $K_\text{eff}$ slightly increases when a non-zero $K_3$ term is considered.

As an additional thickness-dependent magneto-elastic term ($K_\text{elastic}(t_\text{eff})$) can account for the non-linear thickness dependence of $K_\text{eff}t_\text{eff}$, we assume that the electric field $E$ can also modulate $K_\text{elastic}$, notably via the effective interfacial magneto-elastic term $B_\text{i}$. Eq. (3) should then be rewritten as:

$$\xi = \frac{\Delta K_\text{eff} t_\text{eff}}{E} = \frac{\Delta K_\text{i}'}{E} + \frac{1}{t_\text{eff}} \frac{\Delta K_3}{E} \quad (7)$$

where $\frac{\Delta K_i\prime}{E} = \frac{\Delta K_i}{E} + \epsilon_0 \frac{\Delta B_i}{E}$ and $\frac{\Delta K_3}{E} = \gamma \frac{\Delta B_i}{E}$. The data presented in Figure **3** are fitted using Eq. (7) with $\frac{\Delta K_i\prime}{E}$ and $\frac{\Delta K_3}{E}$ as the fitting parameters. The fitting results are summarized in Table 1. The sign difference of $\frac{\Delta K_i\prime}{E}$ and $\frac{\Delta K_3}{E}$ suggests that the voltage-induced contribution of the latter competes with the former. A positive $V_g$ *increases* $H_k$ as well as $K_{eff}t_{eff}$ through $\frac{\Delta K_3}{E}$. With increasing $t_{eff}$, the contribution of $\frac{\Delta K_3}{E}$ on $\xi$ decays and $\xi$ approaches $\frac{\Delta K_i\prime}{E}$ asymptotically for sufficiently high $t_{eff}$ (dashed lines in Figure 3).

These results highlight the importance of including the phenomenological term ($K_3$) in Eqs. (6) and (7) to account for the CoFeB thickness dependence of $K_{eff}t_{eff}$ and $\xi$. With regard to the origin of $\frac{\Delta K_3}{E}$, one may assume that the electronic structure of the CoFeB/MgO interface varies as the average strain of the CoFeB layer changes with its thickness,[19] leading to changes in $\xi$. Alternatively, Naik *et al.* have reported that the electric field can induce strain in the heterostructure through the piezoelectricity of MgO,[21] thereby causing the VCMA via the magnetostriction of CoFeB. If we were to account for the thickness dependence of $\xi$ using the values of the piezoelectricity and magnetostriction provided in Refs. [21] and [39], respectively, we find that neither the sign of $\xi$ nor its dependence on $t_{eff}$ matches with our data (see supplementary information). Moreover, we note that the piezoelectricity mechanism predicts, in CoFeB/MgO/CoFeB junctions, *simultaneous* increase or decrease of the PMA of the bottom and top CoFeB layersCoFeB due to an applied voltage, which is in contrast to what has been found experimentally.[14]

The fact that similar thickness-dependent $\xi$ has not been observed in epitaxial Fe/MgO/Fe (Ref. [37]) or polycrystalline Pt/Co/MgO heterostructures[42] seems to suggest its close relationship to the unique nature of the annealed CoFeB/MgO system, for which the non-linearity in $K_{eff}t_{eff}$ versus $t_{eff}$ appears together with the recrystallization of CoFeB and the onset of the PMA.[39,41] Due to the large lattice mismatch of ~4% between the bulk bcc CoFe and the rock-salt MgO, significant strain is expected to develop at the crystallized interface. Recent *ab initio* calculations revealed that the strain can drastically modify the electronic structure and consequently the VCMA in Ta/FeCo/MgO junctions.[19] However, systematic experimental investigations of the strain development and its implication to VCMA are challenging since it depends on the thickness, the details of each layer constituting the heterostructure, the degree of crystallization at the interface, the boron diffusion profile and many other inter-correlated factors. Here, we introduce a simple $K_3$ term in order to quantitatively describe its contribution to both the MAE and $\xi$. We believe the fact that the VCMA efficiency appears to be highly sensitive to minor details of the structure or conditions may be partially attributed to the negligence of this poorly-understood, thickness-dependent $K_3$ term that may compete with the common interfacial PMA.

In summary, we have experimentally demonstrated that the voltage-induced magnetic anisotropy in heavy metal/CoFeB/MgO heterostructures strongly depends on the effective CoFeB thickness. We find that a higher-order phenomenological term that accounts for the reduction of the areal magnetic anisotropy for thin CoFeB can also describe the change in the areal VCMA efficiency with the CoFeB thickness. The higher order term competes with the conventional interfacial VCMA term, thus reducing the efficiency as the CoFeB thickness is reduced. We infer that a built-in strain in the system, likely at the CoFeB/MgO interface, is responsible for the appearance of the higher order term.

**Supplementary Material**

See supplementary material for the details of the evaluation on the piezo-electric effect on the VCMA efficiency.

**Acknowledgements**

We would like to thank T. Koyama for technical support. This work was partly supported by JSPS Grant-in-Aids for Specially Promoted Research (15H05702), Scientific Research (S)(25220604), ImPACT Program of Council for Science, Technology and Innovation and Spintronics Research Network of Japan.

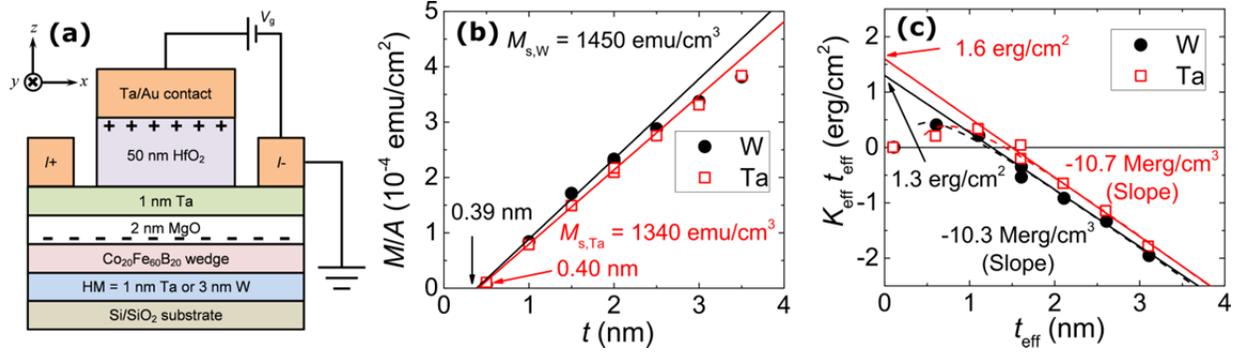

Figure 1: (a) Cross sectional schematic of a Hall bar device with the top $HfO_2$ gate. The heavy metal (HM) underlayer is 1 nm Ta or 3 nm W. A positive gate voltage $V_g$ results in electron accumulation at the CoFeB/MgO interface. $I+$ ($I-$) denotes positive (negative) lead of the probe current. (b) Magnetic moment per area $M/A$ against CoFeB nominal thickness $t$. Black solid circles and open red squares represent, respectively, samples grown on W and Ta underlayers. The magnetization $M_s$ and the magnetic dead layer thickness $t_{DL}$ are given by the slope and the $x$-intercept of the linear fit. (c) Areal effective magnetic anisotropy energy $K_{eff}t_{eff}$ versus effective CoFeB thickness $t_{eff}$. Note that for $t_{eff} < \sim 1$ nm, experimental data deviate significantly from the ideal linear behavior. Linear fits with Eq. 1 (solid lines) excluding these points yield the bulk contribution $K_B - 2\pi M_s$ from the slope and the interfacial anisotropy $K_i$ from the $y$-intercept. Fits using Eq. 6 are shown by the dash line.

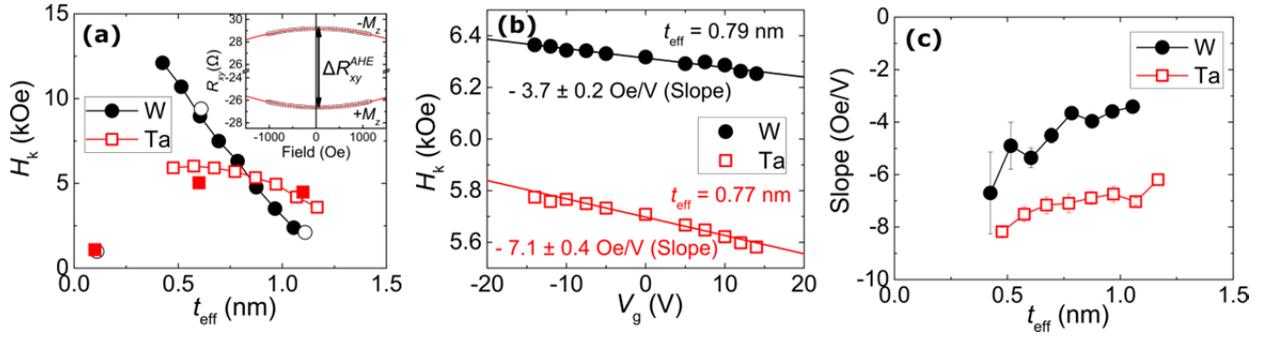

Figure 2: (a) $t_{eff}$ dependence of the anisotropy field $H_k$ of the CoFeB layer. Black solid circles (W underlayer) and red open squares (Ta underlayer) are extracted from quadratic fits of anomalous Hall effect (AHE) data at $V_g$ = 0 V. $H_k$ obtained from vibrating sample magnetometry are plotted in black open circles and red solid squares for W and Ta underlayer films, respectively. Inset shows typical AHE resistance as a function of in-plane field: the definition of $\Delta R_{xy}^{AHE}$ is illustrated. (b) $H_k$ versus $V_g$ comparing two samples having similar $t_{eff}$ and $H_k$ but with different underlayers (W and Ta). Linear fits are performed to evaluate the efficiency of the voltage controlled $H_k$ modulation. (c) Slope of $H_k$ modulation per $V_g$ as a function of $t_{eff}$.

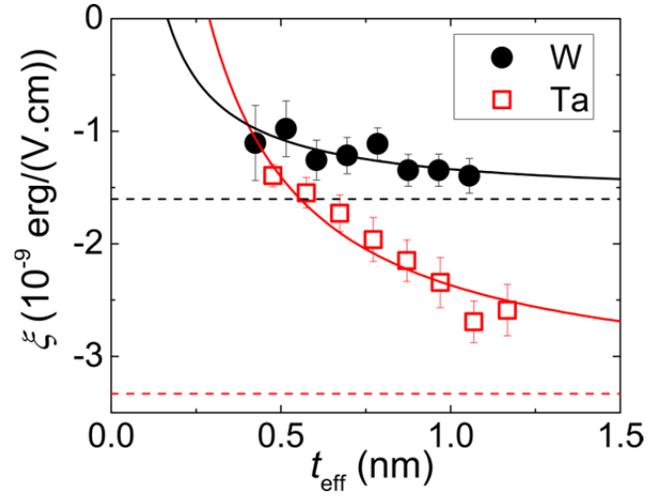

Figure 3: Efficiency of voltage-controlled magnetic anisotropy $\xi$ as a function of $t_{\text{eff}}$. Solid lines are best fits to the experimental data using Eq. 7. The dashed lines represent the conventional interfacial contributions of voltage-controlled magnetic anisotropy $\frac{\Delta K_i'}{E}$ obtained from Eq. (7).

Table 1: Summary of the fitting parameters for the data in Figure 1(c) and Figure 3.

| Seed layer | Néel linear model (Eq. 1) | | Model including $K_3$ term (Eq. 6 & 7) | | | | |
|---|---|---|---|---|---|---|---|
| | $K_B - 2\pi M_s^2$ (erg/cm$^3$) | $K_i$ (erg/cm$^2$) | $K_B' - 2\pi M_s^2$ (erg/cm$^3$) | $K_i'$ (erg/cm$^2$) | $\Delta K_i'/E$ (erg/(V·cm)) | $K_3$ (erg/cm) | $\Delta K_3/E$ (erg/V) |
| Ta | -10.7 x 10$^6$ | 1.59 | -11.5 x 10$^6$ | 2.05 | -3.3 x 10$^{-9}$ | -6.0 x 10$^{-8}$ | 9.6 x 10$^{-17}$ |
| W | -10.3 x 10$^6$ | 1.28 | -10.9 x 10$^6$ | 1.55 | -1.6 x 10$^{-9}$ | -2.8 x 10$^{-8}$ | 2.7 x 10$^{-17}$ |

# Supplementary information of

# Electric field modulation of the non-linear areal magnetic anisotropy energy

Yong-Chang Lau[1,2], Peng Sheng[1], Seiji Mitani[1], Daichi Chiba[3] and Masamitsu Hayashi[1,2]

[1]*National Institute for Materials Science, Tsukuba 305-0047, Japan*

[2]*Department of Physics, The University of Tokyo, Bunkyo, Tokyo 113-0033, Japan*

[3]*Department of Applied Physics, The University of Tokyo, Bunkyo, Tokyo 113-0033, Japan*

**S1. VCMA contribution due to the piezoelectricity of MgO**

According to Naik *et al.*[1], an applied positive bias stretches the MgO along *z* and compresses the MgO evenly along *x* and *y*, which increases the PMA of the top CoFeB free layer. Since the layers are very thin, we assume that similar in-plane biaxial compressive strain will be fully transferred to the bottom CoFeB as well. Using an elongation of 37 pm/V for a 5 nm MgO barrier[1], and with a Poisson ratio of 0.33, Naik *et al* reported $\frac{\Delta \epsilon_0}{E}\Big|_{\text{piezo}} \approx$ -1.2×10$^{-9}$ cm/V[1]. (Note that due to the constraint from the substrate, such estimate using a Poisson ratio of 0.33 gives the upper limit of $\frac{\Delta \epsilon_0}{E}\Big|_{\text{piezo}}$.)

According to Gowtham *et al.* [2], the magneto-elastic anisotropy has a general form $K_{\text{elastic}}(t_{\text{eff}}) = B_{\text{eff}}^{\text{biaxial}}(t_{\text{eff}}) \cdot \epsilon(t_{\text{eff}})$, where $B_{\text{eff}}^{\text{biaxial}}$ is the effective magneto-elastic coupling coefficient and $\epsilon$ is the average in-plane biaxial strain. The parameters are defined as:

$$B_{\text{eff}}^{\text{biaxial}}(t_{\text{eff}}) \approx B_B + B_i/t_{\text{eff}}$$

$$\epsilon(t_{\text{eff}}) \approx \epsilon_0 + \gamma/t_{\text{eff}}$$

$B_B$ and $B_i$ ($\epsilon_0$ and $\gamma$) are the bulk and interfacial contributions to the effective magneto-elastic coupling coefficient (strain). For 1.1 nm < $t_{\text{eff}}$ < 2.0 nm, Gowtham *et al.* obtained:

$$B_B \approx -3.0 \times 10^8 \text{erg/cm}^3$$

$$B_i \approx +24.2 \text{ erg/cm}^2$$

An interpretation of the observed VCMA entirely in terms of the piezoelectricity of MgO would imply:

$$\xi_{\text{piezo}} \equiv \frac{\Delta K_{\text{piezo}}}{E} \approx (B_B t_{\text{eff}} + B_i) \cdot \frac{\Delta \epsilon_0}{E}\bigg|_{\text{piezo}}.$$

As shown in Figure S1, we find that neither the sign of $\xi_{\text{piezo}}$ (>0) nor its trend ($\frac{\partial \xi_{\text{piezo}}}{\partial t_{\text{eff}}} = B_B \cdot \frac{\Delta \epsilon_0}{E} > 0$) matches with our data when $t_{\text{eff}} > 1$ nm.

We note that upon extrapolating to lower thicknesses, $\xi_{\text{piezo}}$ changes sign at $t_{\text{eff}} \sim 0.8$ nm. This is due to the sign change of $B_{\text{eff}}^{\text{biaxial}}$ at this thickness. We have doubt on the validity of such extrapolation since it would imply that the magneto-elastic anisotropy of the CoFeB also changes sign at this thickness. We therefore focus the discussion in a thickness range in which the parameters are evaluated by Gowtham *et al*.

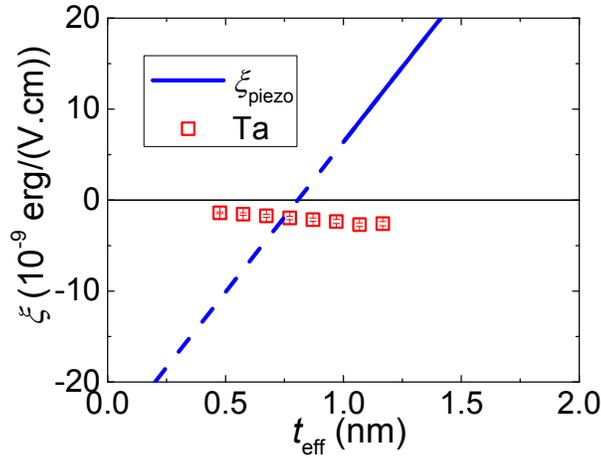

Figure S1: Efficiency of voltage-controlled magnetic anisotropy, $\xi$ as a function of $t_{\text{eff}}$. Red open squares are our experimental data determined from the annealed Ta/CoFeB/MgO heterostructures. Solid line is the expected magneto-elastic contribution via the piezoelectricity $\xi_{\text{piezo}}$. Dashed line is the extrapolated values of $\xi_{\text{piezo}}$ at lower thicknesses.

**Supplementary References:**